\documentstyle [12pt] {article}
\begin{document}
\title {Spinning and Spinning Deviation Equations of Bi-metric Type Theories}
\maketitle

\begin{center}
{\bf {Magd E. Kahil{\footnote{Faculty of Engineering,  Modern Sciences and Arts University, Giza, Egypt  \\
e.mail: mkahil@msa.eun.eg}}} }
\footnote{Egyptian Relativity Group. Cairo, Egypt}
\end{center}

\begin{abstract}Spinning equations of bi-metric types theories of gravity, the counterpart of the Papapetrou spinning equations of motion have been derived as well as their corresponding spinning deviation equations. Due to introducing different types of bi-metric theories, the influence of different curvatures based upon different affine connections , have been examined. A specific Lagrangian function for each type theory has been proposed, in order to derive the set of spinning motions and their corresponding spinning deviation equations.  \\
 
\end{abstract}

\section{Introduction}
 Bi-metric theories of gravity are considered promising theories of gravity in strong fields. These equations have been performed in different stages since last century. In 1940 Rosen introduced such a challenging  gravitational theory, wider than the orthodox general theory of relativity.  The theory is called the bi-metric theory of gravity core of the theory, based on considering any point in the manifold is identified by two reference frames, the first is described in a curved space ; while the second is expressed in a flat space, obtaining its corresponding geodesic equations [1-3].
 Meanwhile, Israelit (1975) [4] solved these equations of motion for test particle, and  Falik and Rosen (1981) extended this study to examine the motion of charged particles. [5] .

Yet, the concept of imposing two metrics, has inspired Moffat [6] to present another version of bi-metric theory of gravity, based on regarding one combined metric produced of these two previous ones. This version of bi-metric theory is considering a variable speed of light (VSL), to be considered  as an alternative solution to dismiss dark energy as mentioned in different theories of gravity [7].

Also, the concept of bi-metric theory has been extended within the context of a modified Newtonian gravity MOND by Milgrom [8]. Owing to its existence as an  alternative remedy to the apparent constant speed of rotation curves in spiral galaxies instead of proposing dark matter particles [9]. As many authors regard dark matter is not centered only within the outer arms of spiral galaxies, but in the core of galaxies subject to strong gravitational fields [10]. So, proposing a novel concept of a bi-metric modified dynamics BIMOND becomes essential to define the behavior of particles that are not responding to the conventional theories of gravity [11].

 Consequently, in 2012 Hassan and Rosen made a paradigm shift in bi-metric theories of gravity [12]. Such a theory is relying on presenting  the concept of  bi-gravity theory, in which there are two types of matter produced by two parallel field equations one is for an ordinary matter, and the other is related to a twin matter, the theory has a vital feature to be considered as a ghost free one [13] . Following this approach, Arkami et al (2014) performed their corresponding path equations [14] .

Moreover , an alternative version of bi-metric theory of gravity has been produced by Verozub based on curing the defect obtained in Einstein's equations for being not invariant under geodesic mappings. In this theory, one may find out that the geodesic equations become invariant in a given coordinate system, i.e. the geodesic mapping act as gauge transformations [15]. Such a tendency  brings forth a theory of gravity able to examine the behavior of trajectories in very strong gravitational fields such as Sgr A* [16]. Also, the theory has been applied on examining the  stability of super-massive objects in strong gravitational fields such as the active galactic nuclei [17].

The aim of our present work is to extend our previous study which was assigned to derive the corresponding path and path deviation equations from of bi-metric type theories of gravity Kahil (2017) [18]. Accordingly, it is essential to find the analogous parts of the Papapetrou equations [19] in bi-metric type theories as similar as  the ones performed in both Einstein-Cartan theories [20] and a specific class of non-Riemannian geometries called absolute parallelism (AP)-geometry [21].

 The reason for studying spinning equations is not only as test particles but also as  extended bodies , in order to study the effect of  their intrinsic properties. This work may help to study in  detail, the stability of these spinning objects orbiting strong gravitational fields, which will be planned in our future work.

The paper is organized as follows: Section 1 shows the transformation from geodesic into a spinning path for short. From Section 2 to Section 8  there are a detailed  derivation for spinning and spinning  deviations for different bi-metric type theories e.g. Rosen's bi-metric the first version theory, Moffat's  variable speed of light, Milgrom's approach of BIMOND, Hassan-Rosen Bigravity theory and  Verozub's bimetric  of geodesic-invariant equations of gravity respectively. Finally, Section 9 presents concluding remarks of the importance of deriving spinning and spinning deviation equations for bi-metric types theories, proposes a road map for extending in future this work  using different geometries apart from the Riemannian one.
\section{Transformation From Test Particles to  Spinning Objects}
Equations of motion for a spinning objects may be derived in twofold , one of them from geodesic equations as being a deviated path from geodesic satisfying the following relation [21]:
\begin{equation}
V^{\alpha} = U^{\alpha} + \beta \frac{D \Psi^{\alpha}}{DS}
 \end{equation}
where $V^{\alpha} = \frac{d x^{\alpha}}{d \tau}$ is the unit tangent vector associated to path $\tau$ and $U^{\alpha} $ is a unit tangent vector of a geodesic defined with the parameter $S$,  $\beta$ is an arbitrary parameter and $\Psi^{\alpha}$ is S-dependent deviation vector associated with one parameter of  a family of geodesics $x^{\mu}(S, \epsilon)$  such that [22]
 $$
 \Psi^{\mu} = \epsilon \frac{\partial x^{\mu}}{\partial \epsilon} |_{\epsilon =0}
 $$

 and $\frac{D}{DS}$ is the covariant derivative with respect to parameter $S$.
i.e.
$$
\frac{D X^{\mu}}{Ds} = \frac{d A^{\mu}}{d S} + \Gamma^{\mu}_{\alpha \beta} A^{\alpha} \frac{dx^{\beta}}{d S},
$$
where, $A^{\mu}$ is an arbitrary vector and $\Gamma^{\mu}_{\alpha \beta}$ is the Chrristoffel symbol. \\
Thus, taking the covariant derivative of (1), and proposing the following relation
\begin{equation}
S^{\alpha \beta} =\sigma(U^{\alpha} \Psi^{\beta}- U^{\beta} \Psi^{\alpha} )
\end{equation}
where $\sigma$ is the magnitude of spin, and $S^{\alpha \beta }$ is the spin tensor, in which $\beta = \frac {\sigma}{m}$, $m$ is the mass of the spinning object.\\ Thus, the geodesic equation becomes,
\begin{equation}
\frac{D U^{\alpha}}{D S} =0,
\end{equation}
and its corresponding geodesic deviation equation turns to be as follows
\begin{equation}
\frac{D^2 \Psi^{\alpha}}{D S^2} = R^{\alpha}_{\rho \sigma \delta} U^{\rho} U^{\sigma} \Psi^{\delta},
\end{equation}
where, $R^{\alpha}_{\beta \rho \sigma}$ is the Riemann curvature tensor, and
$$
\frac{d S }{d \tau} =1.
$$
Consequently , equation (1) reduces, after simple calculations, to the Papapertrou  equation for short [19],
\begin{equation}
 \frac{D V^{\alpha}}{D \tau} =  \frac{1}{2m} R^{\alpha}_{\rho \sigma \delta} S^{\sigma \delta} V^{\rho}
\end{equation}
Accordingly, it can be found that this type of formulation is feasible for deriving spinning objects with no precession with their intrinsic properties.   However, the process of deriving  a generalized method able to obtain the translational and rotational equations need to propose a rival method based  applying the action principle on a  specific Lagrangian as shown in the following part.

\subsection*{The Papapertrou Equation in General Relativity: Lagrangian Formalism}
  It is well known that equation of spinning objects in the presence of gravitational field have been studied extensively. This led us to suggest its corresponding  Lagrangian formalism , using a modified Bazanski Lagrangian [24], for  a spinning and precessing object and their corresponding deviation equation in Riemanian geometry in the following way [21]
\begin{equation}
 L = g_{\alpha \beta} P^{\alpha} \frac{D \Psi^{\beta}}{DS} + S_{\alpha \beta}\ \frac{D \Psi^{\alpha \beta}}{DS}+ \frac{1}{2} R_{\alpha \nu \rho \delta} S^{\rho \delta} U^{\nu}\Psi^{\alpha}+ 2 P_{ [ \alpha}U_{\beta ]} \Psi^{\alpha \beta}
 \end{equation}
 where  $P^{\mu}$ is the momentum vector,
in which,
 $$ P^{\alpha}= m U^{\alpha}+ U_{\beta} \frac{D S^{\alpha \beta}}{DS}.$$
and $\Psi^{\mu \nu}$ is defined as ,S- dependent deviation tensor associated with one parameter of  a family of spin tensor $S^{\mu \nu}(S, \epsilon)$  such that [22]
 $$
 \Psi^{\mu \nu} = \epsilon \frac{\partial S^{\mu \nu}}{\partial \epsilon} |_{\epsilon =0} .
 $$

Applying the Euler-Lagrange equations to get,

\begin{equation}
( \frac{d}{dS} \frac{\partial L}{\partial \dot{\Psi}^{\mu}} - \frac{\partial L}{\partial \Psi^{\mu}})  =0 ,
\end{equation}

and
\begin{equation}
( \frac{d}{dS} \frac{\partial L}{\partial \dot{\Psi}^{\mu \nu}} - \frac{\partial L}{\partial \Psi^{\mu \nu}})  =0 ,
\end{equation}

to obtain the set of spinning for the  spinning object,

\begin{equation}
\frac{DP^{\mu}}{DS}=  \frac{1}{2} R^{\mu}_{\nu \rho \delta} S^{\rho \delta} U^{\nu} ,
\end{equation}
and,
\begin{equation}
\frac{DS^{\mu \nu}}{DS}= 2 P^{ [ \mu}U^{\nu ]}  .
 \end{equation}

Also, applying the following identity on both equations (9)  and (10), to obtain the set of equations may be derived their corresponding deviation equations, using the following identity [25].

\begin{equation}
  A^{\mu}_{; \nu \rho} - A^{\mu}_{; \rho \nu} = R^{\mu}_{\beta \nu \rho} A^{\beta},
 \end{equation}
and
\begin{equation}
  A^{\mu \nu}_{; \nu \rho} - A^{\mu \nu}_{; \rho \nu} =  A^{[ \mu \beta} R^{\nu ]}_{\beta \nu \rho} ,
 \end{equation}

where $A^{\mu}$ and $A^{\mu \nu}$ are both arbitrary vector and tensor respectively . \\
 Multiplying both sides with arbitrary vectors, $U^{\rho} \Psi^{\nu}$ as well as using the following condition . \\
\begin{equation}
 U^{\alpha}_{; \rho} \Psi^{\rho} =  \Psi^{\alpha}_{; \rho } U^{\rho},
\end{equation}
and $\Psi^{\alpha}$ is its deviation vector associated to the  unit vector tangent $U^{\alpha}$.
 Also in a similar way:
\begin{equation}
 S^{\alpha \beta}_{; \rho} \Psi^{\rho} =  \Psi^{\alpha \beta}_{; \rho } U^{\rho},
\end{equation}

 one obtains the corresponding deviation equations [26]
\begin{equation}
\frac{D^2 \Psi^{\mu}}{DS^2}= R^{\mu}_{\nu \rho \sigma}P^{\nu} U^{\rho} \Psi^{\sigma}+  \frac{1}{2} ( R^{\mu}_{\nu \rho \delta} S^{\rho \delta} U^{\nu})_{; \rho} \Psi^{\rho},
\end{equation}
and
\begin{equation}
\frac{D^2\Psi^{\mu \nu}}{DS^2}=   S^{\rho [ \mu} R^{\nu ]}_{\rho \sigma \epsilon} U^{\sigma} \Psi^{\epsilon} + 2 (P^{ [ \mu}U^{\nu ]} )_{; \rho} \Psi^{\rho}.
\end{equation}

\section{Spinning Equation and Spin Deviation of  Rosen's Approach}
In this approach, we are going to derive the corresponding set of spinning objects as an extension to the obtained path and path deviations of Rosen's bimetric theory of gravity [1-3]. These equations are extension to equations of Rosen's geodesic and geodesic deviation using the Bazanski Lagrangian
as derived in [18]. \\
(i){\underline{Case $P^{\alpha} = m U^{\alpha}$}} \\
  In this approach, we are deriving spinning equations for short,  that are obtained from the following Lagrangian:
\begin{equation}
L = ( g_{\alpha \beta}- \gamma_{\alpha \beta} ) U^{\alpha} \frac{\nabla \Psi^{\beta}}{\nabla S}+ S_{\mu \nu} \frac{\nabla \Psi^{\mu \nu}}{\nabla S} + \frac{1}{2m} ( R_{\alpha \beta \gamma \sigma})U^{\alpha}  \Psi^{\beta}S^{\gamma \sigma}
\end{equation}
where  $g_{\mu \nu}$ is the metric tensor of the curved space and $\gamma_{\mu \nu}$ the corresponding metric tensor of the flat space

 where, $ {\frac{\nabla A^{\mu}}{\nabla S}}$ a specific covariant derivative defined as follows [27]:
$$
{\frac{\nabla A^{\mu}}{\nabla S}} = \frac{d A^{\mu}}{d S}+ \Delta^{\mu}_{ \nu \rho} A^{\nu} U^{\rho}
$$
such that
$$
\Delta^{\mu}_{ \nu \rho}= \Gamma^{\mu}_{ \nu \rho}- \hat{\Gamma}^{\mu}_{ \nu \rho}
$$
where  $\hat{\Gamma}^{\mu}_{ \nu \rho}$ is the affine connection of the flat space.
 Using the Bazanski approach [23]  to obtain its path equation by taking the variation with respect to $\Psi^{\alpha}$ and $\Psi^{\alpha \beta}$ respectively.
\begin{equation}
\frac{\nabla U^{\mu}}{\nabla S}= \frac{1}{2m} R^{\alpha}_{. \mu \nu \rho}S^{\nu } U^{\mu},
\end{equation}
and
\begin{equation}
\frac{\nabla S^{\alpha \beta}}{\nabla S} = 0.
\end{equation}
 Thus, applying the law of commutation relations  (11), (12), (13)  and (14) we find their corresponding set of deviation equation to become
\begin{equation}
\frac{\nabla \Psi^{\alpha}}{\nabla S}= R^{\alpha}_{. \mu \nu \rho} U^{\mu} U^{\nu} \Psi^{\rho}   + \frac{1}{2m} {( R^{\alpha}_{. \mu \nu \rho}S^{\rho \nu} U^{\mu} U^{\nu} )} _{; \rho} \Psi^{\rho}
\end{equation}
where  $;$ is the covariant derivative for a curved space. \\
and
\begin{equation}
\frac{\nabla \Psi^{\alpha \beta}}{\nabla S} =    S^{[ \alpha \delta} R^{\beta ]}_{ \delta \mu \rho} U^{\mu}\Psi^{\rho}
\end{equation}

The above set of deviation equation behaves identically as their counterparts in general relativity. \\

(ii) {\underline{Case $P^{\mu} = (m U^{\mu} + U_{\nu} \frac{\nabla S^{\mu \nu}}{\nabla S})$}} \\
In this approach, we are corresponding spinning equations whose momentum is $P^{\mu}$ is describing the case of extending bodies,  to be obtained from the following Lagrangian:
\begin{equation}
L = ( g_{\alpha \beta}- \gamma_{\alpha \beta} ) P^{\alpha} \frac{\nabla \Psi^{\beta}}{\nabla S}+ S_{\mu \nu} \frac{\nabla \Psi^{\mu \nu}}{\nabla S} + \frac{1}{2m} ( R_{\alpha \beta \gamma \sigma})U^{\alpha}  \Psi^{\beta}S^{\gamma \sigma} + 2 P_{[ \mu} U_{\nu ]}\Psi^{\mu \nu}
\end{equation}
where  $g_{\mu \nu}$ is the metric tensor of the curved space and $\gamma_{\mu \nu}$ the corresponding metric tensor of the flat space.

 Thus, we can apply the Bazanski approach to obtain its path equation by taking the variation with respect to $\Psi^{\alpha}$ and $\Psi^{\alpha \beta}$ respectively [24]
\begin{equation}
\frac{\nabla P^{\mu}}{\nabla S}= \frac{1}{2} R^{\alpha}_{. \mu \nu \rho}S^{\rho \nu} U^{\mu} U^{\nu}
\end{equation}
and
\begin{equation}
\frac{\nabla S^{\alpha \beta}}{\nabla S} =  2 P^{[ \alpha} U^{ \beta ]}
\end{equation}

 From the previous equation, we find a new effect of covariant derivative for flat spaces appears even if its associated curvature is zero. This gives the spinning deviation equations are quite different than their counterpart of general relativity [25]

Applying the law of commutation relation as shown in equations (11),(12, (13) and (14),  we find their corresponding set of deviation equation to become

\begin{equation}
\frac{\nabla \Psi^{\alpha}}{\nabla S}= R^{\alpha}_{. \mu \nu \rho} P^{\mu} U^{\nu} \Psi^{\rho}   + \frac{1}{2m} {[( R^{\alpha}_{. \mu \nu \rho}S^{\rho \nu} U^{\mu} U^{\nu} )} _{; \rho} + {( R^{\alpha}_{. \mu \nu \rho}S^{\rho \nu} U^{\mu} U^{\nu} )} _{| \rho} ] \Psi^{\rho}
\end{equation}
and
\begin{equation}
\frac{\nabla \Psi^{\alpha \beta}}{\nabla S} =  S^{[ \alpha \delta} R^{\beta ]}_{ \delta \mu \rho} U^{\mu}\Psi^{\rho} + 2 ( P^{[ \alpha} U^{ \beta ]}_{; \rho} +  P^{[ \alpha} U^{ \beta ]}_{| \rho }\Psi^{\rho} ).
\end{equation}

where  $|$ is the associated covariant derivative flat space. \\
Comparing  (20) and (21) with and (25) and (26), we find out that the effect of different covariant derivatives appear effective, if the object is regarded its intrinsic  properties on the spinning deviation equations.

\section{Spin  and Spin Deviation Equations of Moffat's Approach}
Moffat [6] presented the  framework of variable speed of light VSL satisfying bimetric theory and its causality to reveal the problem of dark energy due to VSL by introducing such a metric in the following way [28].
\begin{equation}
\hat{g}_{\mu \nu} = g_{\mu \nu}  + B \partial_{\mu}\phi \partial_{\nu}\phi
\end{equation}
 While the inverse metrics $g^{\mu \nu} $ satisfies that
\begin{equation}
\hat{g}^{\mu \nu} = g^{\mu \nu}  + \frac{B}{K} \phi^{; \mu} \phi^{; \nu} + KB \sqrt{T_{\mu \nu}},
\end{equation}
where  $ \hat{g}_{\mu \nu}$ defines a specific matter metric tensor of a given matter field, $B$ , $\phi$ is a bi-scalar field , $K$ is an arbitrary constant and $T_{\mu \nu}$ is a given energy-momentum tensor [29]. The corresponding path and path deviation equations for a test particle were obtained using a modified  Bazanski Lagrangian [18].  Accordingly, we suggest its corresponding  the following Lagrangians  to obtain the different  spinning and spinning deviation  equations as follows:

{i} {\underline{Case $ P^{\mu} = m U^{\mu} $}}  \\
\begin{equation}
\hat{L}= \hat{g}_{\mu \nu} U^{\mu} \frac{\hat{D} \Psi^{\nu}}{\hat{D}S}+ {S}_{\mu \nu} \frac{\hat{D} \Psi^{\nu \nu}}{\hat{D}S}  + \frac{1}{2m} \hat{R}_{\alpha \beta \gamma \delta } S^{\gamma \delta} U^{\beta} \Psi^{\alpha}
\end{equation}
where
\begin{equation}
\hat{R}^{\alpha}_{\beta \gamma \delta }= \hat{\Gamma}^{\alpha}_{\beta \delta , \gamma}- \hat{\Gamma}^{\alpha}_{\beta \gamma , \delta } + \hat{\Gamma}^{\nu}_{\beta \delta }\hat{\Gamma}^{\alpha}_{\nu \gamma }-\hat{\Gamma}^{\nu}_{\beta \gamma }\hat{\Gamma}^{\alpha}_{\nu \delta}.
\end{equation}
Taking the variation with respect to $\Psi^{\alpha}$ and $\Psi^{\alpha \beta }$ on (29) to become
\begin{equation}
\frac{\hat{D}U^{\alpha}}{\hat{D}S} = \frac{1}{2m} \hat{R}^{\alpha}_{\beta \gamma \delta } S^{\gamma \delta} U^{\beta},
\end{equation}
and
\begin{equation}
\frac{\hat{D}S^{\alpha \beta}}{\hat{D}S} = 0.
\end{equation}

Similarly, we can obtain their corresponding deviation equation, using the commutation relations as shown in the (11), (12),(13) and (14) to get

\begin{equation}
\frac{\hat{D}^2 \Psi^{\alpha}}{\hat{D}S^2} = \hat{R}^{\alpha}_{\beta \gamma \delta } U^{\gamma} U^{\beta} \Psi^{\delta} + \frac{1}{2m} ( \hat{R}^{\alpha}_{\beta \gamma \delta } S^{\gamma \delta} U^{\beta})_{;\rho} \Psi^{\rho},
\end{equation}
and
 \begin{equation}
\frac{\hat{D}^2 \Psi^{\mu \nu}}{\hat{D}S^2} = S^{ [ \alpha \delta} \hat{R}^{\beta ]}_{\delta \gamma \rho } U^{\gamma} \Psi^{\rho}.
\end{equation}
{ii} {\underline{Case $\hat{P}^{\mu} = (m U^{\mu} + U_{\nu} \frac{\hat{D} S^{\mu \nu}}{\hat{D} S})$}}  \\

\begin{equation}
\hat{L} =  \hat{g}_{\alpha \beta} \hat{P}^{\alpha} \frac{\hat{D} \Psi^{\beta}}{\hat{D} S} + S_{\mu \nu} \frac{ \hat{D} \Psi^{\mu \nu}}{ \hat{D} S} + \frac{1}{2} \hat{R}_{\alpha \beta \gamma \sigma}U^{\alpha}  \Psi^{\beta}S^{\gamma \sigma} + 2P_{[{\mu}} U_{{\nu}]} \Psi^{\mu \nu}
\end{equation}

Thus, taking the variation  with respect to $ \Psi^{\delta}$ and $\Psi^{\delta \sigma}$  respectively to get
\begin{equation}
\frac{\hat{D}\hat{P}^{\alpha}}{\hat{D} S}= \frac{1}{2}\hat{R}^{\alpha}_{. \mu \nu \rho}S^{\rho \nu} U^{\mu} U^{\nu},
\end{equation}
and
\begin{equation}
\frac{ \hat{D} S^{\alpha \beta}}{ \hat{D} S}= 2 P^{[ \alpha} U^{\beta ]}.
\end{equation}
In this case, it can be found that the problem of obtaining the corresponding deviation equation using the commutation relations as shown in the (11), (12),(13) and (14)1  :
 \begin{equation}
\frac{\hat{D}^{2}\Psi^{\alpha }}{ \hat{D} S^{2}}= \hat{R}{\alpha}_{.\mu \nu\rho}\hat{P}^{\mu}U^{\nu}\Psi^{\rho}  +\frac{1}{2} (\hat{R}^{\alpha}_{. \mu \nu \rho}  S^{\nu \rho} U^{\mu})_{; \sigma}{ \Psi^{\sigma}}
\end{equation}
and

\begin{equation}
\frac{ \hat{D}^2 \Psi^{\alpha \beta}}{ \hat{D} S^2 }=  (S^{[ \alpha \delta} \hat{R}^{\beta ]}_{ \mu \nu \delta} ) \Psi^{\mu} U^{\nu} + 2 (\hat{P}^{[ \alpha} U^{\beta ]})_{; \rho} \Psi^{\rho}
\end{equation}

The above equations are similar to their counterparts of general relativity due to combining the two metric tensor into one metric.

\section{Spin and Spin deviation Equations of BIMOND Type Theories}
   Modified Newtonian gravity paradigm (MOND) has been introduced by Milgram to reveal the discrepancies found in rotation curves of spiral galaxies [9]. He introduced a constant $a_{0}$ of acceleration units to regulate the transition between Newtonian dynamics and General Relativity. Such a constant has a similar effect as $\hbar$ in quantum mechanics and $G$ the gravitational constant in theories of gravity. It has been found that $a_{0}= \frac{c}{2 \pi H_{0}}$ where $c$ is the speed of light and $H_{0}$ is Hubble constant [11]. \\
Thus, in the context of bi-metric theories, Milgram has extended its significance to embody bi-metric theories of gravity. In that sense, there are two metrics $g_{\mu \nu}$ is responsible for describing the ordinary matter, and ${\gamma_{\mu \nu}}$ is proposed to express twin matter. The difference between their affine connection is regulated by a tensor $C^{\alpha}_{\beta \gamma}$.
 \begin{equation}
C^{\alpha}_{\beta \rho} = \Gamma^{\alpha}_{\beta \rho}- \bar\Gamma^{\alpha}_{\beta \rho},
\end{equation}
such that
$$
g_{\mu \nu ; \rho} = g_{\delta \nu} C^{\delta}_{\mu \rho}  +  g_{\delta \mu} C^{\delta}_{\nu \rho},
$$
and
$$
\gamma_{\mu \nu | \rho} = - \gamma_{\delta \nu} C^{\delta}_{\mu \rho}  -  \gamma_{\delta \mu} C^{\delta}_{\nu \rho}.
$$

Accordingly, $C^{\alpha}_{\beta \gamma}$ may be connected with $a_{0}$ to produce a quantity able to switch from the limits of GR at $a_{0} \rightarrow  0$ and   the MOND limit $a_{0} \rightarrow  \infty$ . \\
The advantage of BIMOND is playing the role to measure the gravitational lensing in an accurate way. It also has an impact to examine the behavior of galactic dark matter, dark matter and dark energy. This gives rise to regard BIMOND a gravitational theory able to study gravity in strong fields such as the core of black holes [12].

 As we obtained previously, the path equations of test particles using BIMOND theory [18]. Thus, it is mandatory to extend this study to examine the behavior of spinning objects in this situation.\\
 From this perspective, we are going to derive the  relevant equations for spinning objects in the presence of BIMOND
by suggesting the following Lagrangians for the following cases:

i. $P^{\alpha} = m U^{\alpha}$
\begin{equation}
L=  \hat{g}_{\mu \nu} U^{\mu} \frac{\bar{D}^2 \Psi^{\nu}}{DS^2} + S_{\mu \nu}\frac{\bar{D}^2 \Psi^{\mu \nu}}{DS^2}  +\frac{1}{2m}( R_{\alpha. \mu \nu \rho}- \bar{R}_{\alpha. \mu \nu \rho}) S^{\nu \rho} U^{\mu} \Psi^{\alpha}.
\end{equation}
Taking the variation with respect $\Psi^{\alpha}$ and $\Psi^{\alpha  \beta}$ we obtain

\begin{equation}
 \frac{\bar{D}U^{\alpha}}{\bar{D} S}= \frac{1}{2m} ( R^{\alpha}_{. \mu \nu \rho}- \bar{R}^{\alpha}_{. \mu \nu \rho}) S^{\rho \nu} U^{\mu}
\end{equation}

where
$$
\bar{R}^{\alpha}_{\beta \gamma \delta }= \bar{\Gamma}^{\alpha}_{\beta \delta , \gamma}- \bar{\Gamma}^{\alpha}_{\beta \gamma , \delta } + \bar{\Gamma}^{\nu}_{\beta \delta }\bar{\Gamma}^{\alpha}_{\nu \gamma }-\hat{\Gamma}^{\nu}_{\beta \gamma }\bar{\Gamma}^{\alpha}_{\nu \delta}.
$$

\begin{equation}
 \frac{\bar{D} S^{\alpha \beta}}{\bar{D}S}=0
\end{equation}

While their corresponding set of deviation equations  may be derived using the commutation relations (11), (12),  (13) and (14), to become

\begin{equation}
 \frac{\bar{D^2}\Psi^{\alpha}}{\bar{D}S^2}= ( R^{\alpha}_{. \mu \nu \rho}- \bar{R}^{\alpha}_{. \mu \nu \rho}) U^{\mu} U^{\nu} \Psi^{\rho} +  \frac{1}{2m} ( [ R^{\alpha}_{. \mu \nu \rho} S^{\rho \nu} U^{\mu} ]_{; \delta} -  [ \bar{R}^{\alpha}_{. \mu \nu \rho} S^{\rho \nu} U^{\mu} ]_{| \delta}  ) \Psi^{\delta}
\end{equation}

and

\begin{equation}
 \frac{\bar{D^2}\Psi^{\alpha \beta}}{\bar{D}S^2}=  ( S^{[ \alpha \delta }R^{\beta ]}_{\delta  \sigma \rho} - S^{[ \alpha \delta }\bar{R}^{\beta ]}_{\delta  \sigma \rho} ) U^{\sigma} \Psi^{\rho}
 \end{equation}

ii. {\underline{Case $\bar{P}^{\mu} = (m U^{\mu} + U_{\nu} \frac{\bar{D} S^{\mu \nu}}{\bar{D} S})$}} \\
\begin{equation}
L=  \hat{g}_{\mu \nu} \bar{P}^{\mu} \frac{\bar{D}^2 \Psi^{\nu}}{DS^2} + S_{\mu \nu}\frac{\bar{D}^2 \Psi^{\mu \nu}}{DS^2}  +\frac{1}{2}( R_{\alpha. \mu \nu \rho}- \bar{R}_{\alpha. \mu \nu \rho}) S^{\nu \rho} U^{\mu} \Psi^{\alpha} + 2 \bar{P}_{[ \mu} U_{\nu ]} \Psi^{\mu \nu}
\end{equation}
Taking the variation with respect $\Psi^{\alpha}$ and $\Psi^{\alpha  \beta}$ we obtain

\begin{equation}
 \frac{\bar{D}U^{\alpha}}{\bar{D} S}= \frac{1}{2} ( R^{\alpha}_{. \mu \nu \rho}- \bar{R}^{\alpha}_{. \mu \nu \rho}) S^{\rho \nu} U^{\mu}
\end{equation}

and

\begin{equation}
 \frac{\bar{D} S^{\alpha \beta}}{\bar{D}S}= 2 \bar{P}^{[ \alpha} U_{\beta ]},
\end{equation}

while the set of deviation equations may be derived using the commutation relations as expressed in (11), (12),(13) and (14) to become

\begin{equation}
 \frac{\bar{D^2}\Psi^{\alpha}}{\bar{D}S^2}= ( R^{\alpha}_{. \mu \nu \rho}- \bar{R}^{\alpha}_{. \mu \nu \rho}) \bar{P}^{\mu} U^{\nu} \Psi^{\rho} +  \frac{1}{2} ( [ R^{\alpha}_{. \mu \nu \rho} S^{\rho \nu} U^{\mu} ]_{; \delta} -  [ \bar{R}^{\alpha}_{. \mu \nu \rho} S^{\rho \nu} U^{\mu} ]_{| \delta}  ) \Psi^{\delta},
\end{equation}
and
\begin{equation}
\frac{\bar{D^2}\Psi^{\alpha \beta}}{\bar{D}S^2}=   ( S^{[ \alpha \delta }R^{\beta ]}_{\delta  \sigma \rho} - S^{[ \alpha \delta }\bar{R}^{\beta ]}_{\delta  \sigma \rho} ) U^{\sigma} \Psi^{\rho} + 2 ( \bar{P}^{[ \alpha} U^{\beta ]})_{; \rho} \Psi^{\rho} +    2 ( \bar{P}^{[ \alpha} U^{\beta ]})_{| \rho} \Psi^{\rho}.
 \end{equation}
Equations (49) and (50) have shown that the effect of two different covariant derivatives is regarded for an object regarding its intrinsic properties . Such a relationship makes, the bi-metric theory different the conventional  general relativity.

\section{Generalized Spin  and Spin Deviation Equations of Bi-metric Theories}
 Hossenfelder [30] has introduced an alternative version of bi-metric theory, having two different metrics ${g}$ and ${h}$ of Lorentzian signature on a manifold ${M}$ one is defined in tangential space TM and the other is in its co-tangential space T*M respectively. These can be regarded as two sorts of  matter and twin matter, existing individually , each of them has its own field equations as defined within Riemannian geometry.
\begin{equation}
 dS^2 = g_{\mu \nu}dx^{\mu} dx^{\nu},
\end{equation}
 and
\begin{equation}
d\tau^2 = h_{\mu \nu}dx^{\mu} dx^{\nu}.
\end{equation}.
 Thus, as a tendency to derive the spinning and spinning equations as an extension to the previous work in [18].  We  suggest a Lagrangian able to describe two independent sets of a generalized spinning and spinning deviation equations, after applying a specific  action principle,  with taking into considerations new additive terms : twin matter $\bar{m}$,the twin momentum $\bar{P}^{\alpha}$, The twin unit tangent vector $V^{\alpha} $ ,the twin deviation a vector $\Phi^{\alpha}$ , the twin spinning tensor $\bar{S}^{\alpha \beta} $ and the spinning deviation tensor $\Phi^{\alpha \beta}$, provided that $ \frac{d \tau}{dS} =0$ \\ \\
i. Case $P^{\mu}= m U^{\mu}$ and $\bar{P}^{\mu}= \bar{m} V^{\mu}$
\begin{equation}
L= g_{\mu \nu} U^{\mu} \Psi_{; \nu} U^{\nu} + h_{\mu \nu} V^{\mu} \Phi_{| \nu} V^{\mu} + S_{\mu \nu} \Psi^{\mu \nu}_{; \rho} U^{\rho} + \bar{S}_{\mu \nu} \Phi^{\mu \nu}_{| \rho} V^{\rho} + \frac{1}{2m} R_{\alpha \beta \gamma \delta} U^{\beta} S^{\gamma \delta} \Psi^{\alpha} + \frac{1}{2\bar{m}} S_{\alpha \beta \gamma \delta} V^{\beta} \bar{S}^{\gamma \delta}\Phi^{\alpha} ,
\end{equation}
where $S^{\mu}_{\nu \rho \sigma}$ is the associated curvature obtained using the matric of twin matter $h_{\mu \nu}$.

Taking the variation with respect to $\Psi^{\alpha}$ and $ \Phi^{\alpha}$ to get
\begin{equation}
\frac{DU^{\alpha}}{DS}= \frac{1}{2m} R^{\alpha}_{\beta \gamma \delta} U^{\beta} S^{\gamma \delta} ,
\end{equation}
and
\begin{equation}
\frac{\bar{D}V^{\alpha}}{\bar{D} \tau}= \frac{1}{2\bar{m}} S^{\alpha}_{\beta \gamma \delta} V^{\beta} \bar{S}^{\gamma \delta}.
\end{equation}

Also, taking the variation with respect to $\Psi^{\alpha \beta}$ and $ \Phi^{\alpha \beta}$ to get
\begin{equation}
\frac{DS^{\alpha \beta}}{DS}= 0 ,
\end{equation}
and
\begin{equation}
\frac{D\bar{S}^{\alpha}}{\bar{D} \tau}= 0.
\end{equation}
While their corresponding Spin deviation equations are obtained using the commutation relations (11),(12, (13) and (14) to become:
\begin{equation}
\frac{D^2\Psi^{\alpha}}{DS^2}= R^{\alpha}_{\beta \gamma \delta} U^{\gamma} U^{\beta} \Psi^{\delta} + \frac{1}{2m}( R^{\alpha}_{\beta \gamma \delta} U^{\beta} S^{\gamma \delta})_{; \delta} \Psi^{\delta} ,
\end{equation}
and
\begin{equation}
\frac{\bar{D}^2\Phi^{\alpha}}{\bar{D}\tau^2}= S^{\alpha}_{\beta \gamma \delta} V^{\gamma} V^{\beta} \Phi^{\delta} +  \frac{1}{2\bar{m}}(  S^{\alpha}_{\beta \gamma \delta} V^{\beta} \bar{S}^{\gamma \delta} )_{| \delta} \Phi^{\delta} ,
\end{equation}
as well as
\begin{equation}
\frac{D^2\Psi^{\alpha \beta}}{DS^2}=  S^{\alpha \sigma}R^{\beta}_{\sigma \gamma \delta} U^{\gamma}  \Psi^{\delta}  ,
\end{equation}
and
\begin{equation}
\frac{\bar{D}^2\Phi^{\alpha \beta}}{\bar{D}\tau^2}= \bar{S}^{\alpha \sigma}S^{\beta}_{\sigma \gamma \delta} V^{\gamma}  \Phi^{\delta} ,
\end{equation}

ii. Case $P^{\mu} = m U^{\mu}+ U_{\nu}\frac{S^{\mu \nu}}{D S } $ and $\bar{P}^{\mu} =  \bar{m} V^{\mu}+ V_{\nu} \frac{\bar{S}^{\mu \nu}}{D S } $
\begin{equation}
L= g_{\mu \nu} P^{\mu} \Psi_{; \nu} U^{\nu} + h_{\mu \nu} \bar{P}^{\mu} \Phi_{| \nu} V^{\mu} + S_{\mu \nu} \Psi^{\mu \nu}_{; \rho} U^{\rho} + \bar{S}_{\mu \nu} \Phi^{\mu \nu}_{| \rho} V^{\rho} + f_{\alpha} \Psi^{\alpha} + \bar{f_{\mu}}\Phi^{\alpha} + 2 P_{[ \mu} U^{\nu ]}\Psi^{\mu \nu} + 2 \bar{P}_{[ \mu} V^{\nu ]}\Phi^{\mu \nu} ,
\end{equation}

where $ f_{\mu} =  \frac{1}{2} R_{\mu \beta \gamma \delta} U^{\beta} S^{\gamma \delta} $ and $ \bar{f}_{\mu} = \frac{1}{2} S_{\mu \beta \gamma \delta} V^{\beta} \bar{S}^{\gamma \delta} $

Taking the variation with respect to $\Psi^{\alpha}$ and $ \Phi^{\alpha}$ to get
\begin{equation}
\frac{DU^{\alpha}}{DS}= \frac{1}{2m} R^{\alpha}_{\beta \gamma \delta} U^{\beta} S^{\gamma \delta} ,
\end{equation}
and
\begin{equation}
\frac{\bar{D}V^{\alpha}}{\bar{D} \tau}= \frac{1}{2\bar{m}} S^{\alpha}_{\beta \gamma \delta} V^{\beta} \bar{S}^{\gamma \delta}.
\end{equation}

Also, taking the variation with respect to $\Psi^{\alpha \beta}$ and $ \Phi^{\alpha \beta}$ to get
\begin{equation}
\frac{DS^{\alpha \beta}}{DS}= 0 ,
\end{equation}
and
\begin{equation}
\frac{D\bar{S}^{\alpha}}{\bar{D} \tau}= 0.
\end{equation}
Thus, their corresponding spin deviation equations are obtained using the commutation relations to become
\begin{equation}
\frac{D^2\Psi^{\alpha}}{DS^2}= R^{\alpha}_{\beta \gamma \delta} U^{\gamma} U^{\beta} \Psi^{\delta} + \frac{1}{2m}( R^{\alpha}_{\beta \gamma \delta} U^{\beta} S^{\gamma \delta})_{; \delta} \Psi^{\delta} ,
\end{equation}
and
\begin{equation}
\frac{\bar{D}^2\Phi^{\alpha}}{\bar{D}\tau^2}= S^{\alpha}_{\beta \gamma \delta} V^{\gamma} V^{\beta} \Phi^{\delta} +  \frac{1}{2\bar{m}}(  S^{\alpha}_{\beta \gamma \delta} V^{\beta} \bar{S}^{\gamma \delta} )_{| \delta} \Phi^{\delta} ,
\end{equation}
as well as
\begin{equation}
\frac{D^2\Psi^{\alpha \beta }}{DS^2}=  S^{[ \alpha \rho }R^{\beta ]}_{\rho  \gamma \delta} U^{\gamma} \Psi^{\delta} + 2( P^{[ \alpha} U^{\beta ]} )_{; \delta} \Psi^{\delta} ,
\end{equation}
and
\begin{equation}
\frac{\bar{D}^2\Phi^{\alpha \beta }}{\bar{D} \tau^2}=  \bar{S}^{[ \alpha \rho }R^{\beta ]}_{\rho  \gamma \delta} U^{\gamma} \Psi^{\delta} + 2( P^{[ \alpha} U^{\beta ]} )_{; \delta} \Psi^{\delta} .
\end{equation}

\section{Spin and Spin Deviation  Equations of Bi-gravity Type Theories}

Recently, Arkami et al [14] have suggested the two metrics $g_{\mu \nu}$ and $h_{\mu \nu}$  are connected with each other by a specific quasi-metric free from ghost in the following manner  ,

If one considers  the two metrics can be related to each other, they can be combined in one metric as a quasi-metric one [31] such that:
\begin{equation}
\tilde{g}_{\mu \nu} = \alpha^{2}_{g} g_{\mu \nu} + \alpha^{2}_{h}h_{\mu \nu} + \alpha_{g} \alpha_{h} [ \frac{d \tau}{d S}( g_{\mu \nu} - U_{\mu}U_{\nu} ) + \frac{d S}{d \tau} ( h_{\mu \nu} - V_{\mu}V_{\nu}) + 2 U_{( \mu} V_{\nu )} ] ,
\end{equation}
where $\alpha_{g}$ and $\alpha_{h}$  are the coupling strengths, and their corresponding line element becomes
\begin{equation}
dS^{2} =  ( \alpha_{g}^2 g_{\mu \nu} + \alpha_{h}^2 h_{\mu \nu}) dx^{\mu}dx^{\nu} +  2\alpha_{g}\alpha_{f} \sqrt{( g_{\mu \nu} h_{\rho \sigma} dx^{\mu}dx^{\nu}dx^{\rho}dx^{\sigma}   )}.
\end{equation}

However, applying the action principle on the Lagrangian function (53)  to obtain the corresponding spinning equations of bi-gravity  theory is expressed  for the case $P^mu = m U^{\mu} $ and $P^mu = \bar{m} V^{\mu}$in the following way:
\begin{equation}
( \frac{d}{dS} \frac{\partial L}{\partial \dot{\Psi}^{\alpha}} - \frac{\partial L}{\partial \Psi^{\alpha}}) +
( \frac{d \tau}{dS} ) ( \frac{d}{d \tau} \frac{\partial L}{\partial \dot{\Phi}^{\alpha}}- \frac{\partial L}{\partial \Phi^{\alpha}}) =0 ,
\end{equation}

to give the spinning analog whose geodesic-like  has mentioned by Arkani et al (2014)

\begin{equation}
 g_{\mu \nu} ( \frac{D U^{\mu}}{D S} + \frac{1}{2m} R^{\mu}_{\nu \rho \sigma} S^{\rho \sigma} U^{\nu})  + h_{\mu \nu} ( \frac{\bar{D}V^{\mu}}{\bar{D} \tau} + \frac{1}{2 \bar{m}} S^{\mu}_{ \nu \rho \sigma} \bar{S}^{\rho \sigma} V^{\nu} ) \frac{d \tau}{ ds }  =  0
\end{equation}
Yet, extending the same technique of the Bazanski approach, we  obtain its deviation equations to  obtain:
\begin{equation}
 g_{\mu \alpha} {[ \frac{D^2\Psi^{\alpha}}{DS^2}+ R^{\alpha}_{\beta \delta \gamma} U^{\gamma} U^{\beta} \Psi^{\delta} ]} +
{( \frac{d \tau}{d S})}^2  \gamma_{\mu \alpha} {[ \frac{D^2 \Phi^{\alpha} }{D \tau^2}+ R^{\alpha}_{\beta \delta \gamma} V^{\gamma} V^{\beta} \Phi^{\delta} ]}  ,
=0
\end{equation}

Conerquently, applying the action principle on the Lagrangian function (62) to obtain the corresponding spinning equations of bi-gravity  theory is expressed in the following way:
\begin{equation}
( \frac{d}{dS} \frac{\partial L}{\partial \dot{\Psi}^{\alpha \beta}} - \frac{\partial L}{\partial \Psi^{\alpha \beta}}) +
( \frac{d \tau}{dS} ) ( \frac{d}{d \tau} \frac{\partial L}{\partial \dot{\Phi}^{\alpha \beta}}- \frac{\partial L}{\partial \Phi^{\alpha \beta}}) =0 ,
\end{equation}

to give the same an extended  results to what was mentioned by Arkani et al (2014) for spinning objects for short
\begin{equation}
 \alpha_{g} g_{\mu \nu} ( \frac{D S^{\mu \nu}}{D S} +  S^{[ \mu \delta}R^{\nu ]}_{\delta \rho \sigma} U^{\rho} )  + \alpha_{h} h_{\mu \nu} ( \frac{\bar{D}V^{\mu}}{\bar{D} \tau} + \frac{1}{2 m} S^{\mu}_{ \nu \rho \sigma} \bar{S}^{\rho \sigma} V^{\nu} ) \frac{d \tau}{ ds }  =  0
\end{equation}
Applying the same technique of the Bazanski approach, we  obtain its deviation equations
to  obtain:

\begin{equation}
g_{\mu \alpha}g_{\nu \beta} {[ \frac{D^2 \Psi^{\alpha}}{DS^2}+ F^{\alpha \beta}}_{\gamma} U^{\gamma} ] +  h_{\mu \alpha}h_{\nu \beta} {[ \frac{\bar{D}^2 \Psi^{\alpha}}{ \bar{D} {\tau}^2}+ \bar{F}^{\alpha \beta}_{\gamma}  V^{\gamma}} ] = 0
\end{equation}
where $$ F^{\alpha \beta}_{\gamma}= 2 S^{[ \alpha \sigma} R^{\beta ]}_{\beta \gamma \delta} U^{\gamma}  \Psi^{\delta}$$ and $$\bar{F}^{\alpha \beta}_{\gamma} = 2 \bar{S}^{[ \alpha \sigma} S^{\beta ]}_{\beta \gamma \delta} V^{\gamma}  \Phi^{\delta} $$
In a similar process, we can find the following equations for the case $P^{\mu} \neq  m U^{\mu} $ and $P^{\mu} \neq {m} V^{\mu}$

Thus, we get the corresponding spinning equations for precessing objects  as mentioned by Arkani et al (2014),
\begin{equation}
 \alpha_{g}g_{\mu \nu} ( \frac{D P^{\mu}}{D S} +  R^{\mu}_{\nu \rho \sigma} S^{\rho \sigma} U^{\nu})  + \alpha_{h} h_{\mu \nu} ( \frac{\bar{D} \bar{P}^{\mu}}{\bar{D} \tau} + \frac{1}{2 } S^{\mu}_{ \nu \rho \sigma} \bar{S}^{\rho \sigma} V^{\nu} ) \frac{d \tau}{ ds}  =  0.
\end{equation}
Applying the same technique of the Bazanski approach, we  obtain its deviation equations
to  obtain
\begin{equation}
( \frac{d}{dS} \frac{\partial L}{\partial \dot{\Psi}^{\alpha \beta}} - \frac{\partial L}{\partial \Psi^{\alpha \beta}}) +
( \frac{d \tau}{dS} ) ( \frac{d}{d \tau} \frac{\partial L}{\partial \dot{\Phi}^{\alpha \beta}}- \frac{\partial L}{\partial \Phi^{\alpha \beta}}) =0 .
\end{equation}

Applying the same technique of the Bazanski approach, we  obtain its deviation equations to  obtain
\begin{equation}
g_{\mu \alpha}g_{\nu \beta} {[ \frac{D^2\Psi^{\alpha \beta}}{DS^2}+ F^{\alpha \beta}_{\gamma} P^{\gamma} + 2 (P^{[ \alpha}U^{\beta]} )_{; \rho} \Psi^{\rho} ]} + h_{\mu \alpha}h_{\nu \beta} {[ \frac{\bar{D}^2 \Phi^{\alpha}}{\bar{D}{\tau}^2}+ \bar{F}^{\alpha \beta}_{\gamma}\bar{P}^{\gamma} + 2 (\bar{P}^{[ \alpha}V^{\beta ]} )_{| \rho} \Phi^{\rho}]}.
\end{equation}
 From the above set of equations (77) and (81), we find that these equations are different than their counterparts in Riemaiann equation, but if we use the metric as defined in (71), then we obtain the following equations for spinning and spinning deviations for bi-gravity theory as similar to the Moffat version of bi-metric theory.\\
Thus, the  corresponding Lagrangian for $P^{\alpha} = m U^{\alpha}$ may be expressed as

 \begin{equation}
 \tilde{L}= \tilde{g}_{\mu \nu} U^{\mu} ( \frac{d \Psi^{\nu} }{dS} + \tilde{\Gamma}^{\nu}_{\rho \delta} \Psi^{\rho} U^{\delta}  + \frac{1}{2m} \tilde{R}_{\mu \nu \rho \sigma} S^{\rho \sigma}U^{\nu} \Psi^{\mu}.
 \end{equation}
such that, the affine connection may be expressed in terms of $ \tilde{\Gamma}^{\alpha}_{\beta \sigma}$ as defined as

$$
  \tilde{\Gamma}^{\alpha}_{\beta \sigma} =  \frac{1}{2}\tilde{g}^{\alpha \delta}( \tilde{g}_{\sigma \delta ,\beta }  +\tilde{g}_{\delta \beta , \sigma } -\tilde{g}_{\beta \sigma ,\delta} )
$$
provided that its corresponding curvature,
$$
\tilde{R^{\alpha}_{\beta \gamma \sigma}} = \tilde{\Gamma}^{\alpha}_{\beta \sigma , \gamma}-  \tilde{\Gamma}^{\alpha}_{\beta \gamma ,\sigma} +  \tilde{\Gamma}^{\rho}_{\beta \sigma} \tilde{\Gamma}^{\alpha}_{\rho \gamma} - \tilde{\Gamma}^{\rho}_{\beta \gamma} \tilde{\Gamma}^{\alpha}_{\rho \sigma}.
$$
  Taking the variation respect to $\Psi^{\mu}$ to obtain the spinning equation for short in the following way
 \begin{equation}
 \frac{\tilde{D}U^{\alpha}}{\tilde{D}S^{2}}= \frac{1}{2m} \tilde{R}^{\alpha}_{ \nu \rho \sigma} S^{\rho \sigma}U^{\nu} ,
\end{equation}
Consequently, applying the commutation laws (9) and (10) on equation (83) to obtain its corresponding path deviation equation,
 \begin{equation}
 \frac{\tilde{D}^{2}\Psi^{\alpha}}{\tilde{D}S^{2}}= \tilde{R}^{\alpha}_{.\mu \nu\rho}U^{\mu}U^{\nu}\Psi^{\rho} +\frac{1}{2m} ( \tilde{R}^{\alpha}_{ \nu \rho \sigma} S^{\rho \sigma}U^{\nu})_{\sigma}\Psi^{\sigma}  .
 \end{equation}

 Moreover, the spinning and spinning deviation equations with precession become,

\begin{equation}
\tilde{L} = \tilde{g}_{\alpha \beta} U^{\alpha}\frac{\tilde{D} \Psi^{\beta}}{\tilde{D} S} +S_{\alpha \beta} \frac{\tilde{D} \Psi^{ \alpha \beta}}{\tilde{D} S} + f_{\alpha}\Psi^{\alpha} + f_{\alpha \beta}\Psi^{\alpha \beta},
\end{equation}
to give
\begin{equation}
\frac{\tilde{D}U^{\mu}}{\tilde{DS}} = f^{\mu},
\end{equation}
and
\begin{equation}
\frac{\tilde{D}S^{\mu \nu}}{\tilde{DS}} = f^{\mu \nu}.
\end{equation}
Accordingly, its corresponding deviation equation becomes:
\begin{equation}
\frac{\tilde{D}^{2}\Psi^{\alpha}}{\tilde{D}S^{2}}= \tilde{R}^{\alpha}_{.\mu \nu\rho}U^{\mu}U^{\nu}\Psi^{\rho} + f^{\alpha}_{; \rho} \Psi^{\rho}
\end{equation}
and
\begin{equation}
\frac{\tilde{D}^{2}\Psi^{\mu \nu}}{\tilde{D}S^{2}}= S^{\mu \delta} \tilde{R}^{\nu}_{.\delta \sigma \rho}U^{\sigma}U \Psi^{\rho} + f^{\mu \nu}_{; \rho} \Psi^{\rho}.
\end{equation}
The above equations of spinning objects in different types of bi-gravity metrics as two independent metrics and one combined one having a free-ghost particles, may give rise to visualize the effect of different curvatures affecting the proper mass and twin mass as well as representing their accompanied spin and twin spin tensors. The combined metric tensors of bi-gravity as expressed in Riemannian geometry may be expressed geometrically using Finsler geometry. Such a type of work will be assigned in our future work.

\section{Spinning and Spinning Deviation Equations of Bi-metric Invariant-Gravitation Theory }
A rival description of gravity by Verozub has been proposed in his version of bi-metric theory of gravity [14]. It has been regarded that gravity can be described in two different geometries [31], which can be expressed in both Minkowski space, an inertial frame of reference IRF,  and Riemaian space as a co-moving reference frame CRF. From this perspective, any quantity in Riemannian, such as the metric tensor $g_{\mu \nu} (x)$ can be transformed as a function of spin-2 force field space $\psi_{\mu \nu}$ to be expressed as $g_{\mu \nu} (\psi)$. While the ordinary derivatives in Riemannian space are transformed as covariant derivatives in Minkowski space. This led Veozub to regard the Christoffel symbols in Riemannian to be transferred based on the mapping  which can new affine  connections in Minkowski space functions of $\psi_{\mu \nu}$ which may be defined as [16],
\begin{equation}
\bar{\Gamma}^{\alpha}_{\beta \gamma} (\psi) = {\Gamma}^{\alpha}_{\beta \gamma} (x) + \delta^{\alpha}_{\beta} \phi_{\gamma}(x) + \delta^{\alpha}_{\beta} \phi_{\gamma}(x)
\end{equation}
$$
\phi_{\mu} = \frac{1}{2(n+1)} \frac{\partial}{\partial x^{\mu}} ln \frac{\bar{g}}{{g}},
$$
 where $\phi(x)$ are arbitrary differentiable function, defined due to implement bi-metric function  as  a relation between two metrics ${g_{\mu \nu}}$ before and $\bar{g_{\mu \nu}}$.\\
 If one defines its associate Christoffel symbol $$\bar{\Gamma}^{\gamma}_{\alpha \beta} (\psi) = \frac{1}{2} G^{\gamma \sigma}( G_{ \sigma \beta, \alpha} + G_{\alpha \sigma , \beta}  - G_{\alpha \beta , \sigma}) $$

where, $G_{\mu \nu} = g_{\mu \nu} (\psi)$ and its corresponding curvature tensor becomes
\begin{equation}
\bar{R^{\gamma}_{\alpha \beta \delta}} (\psi) = \bar{\Gamma}^{\gamma}_{\alpha \delta , \beta} - \bar{\Gamma}^{\gamma}_{\alpha \beta , \delta} + \bar{\Gamma}^{\rho}_{\alpha \delta}\bar{\Gamma}^{\rho ,\beta} - \bar{\Gamma}^{\rho}_{\alpha \beta}\bar{\Gamma}^{\rho \delta}.
\end{equation}
Accordingly, such a type of description is able to solve problems of strong gravity and stability problems nearby super massive black holes.

Moreover, for a point mass moving in co-moving reference frame (the Riemannian space) may be detected as moving along a geodesic line, while it can be considered as point mass moving under force field is observed from the inertial frame IRF (the Minkowski space).\\
From this perspective, it can be considered that for a point mass moving in an inertial reference frame under the effect of a given field $\psi^{\mu \nu}$, and an observer located in a co-moving reference frame (CRF), may find that the particle is moving on a geodesic line for a space-time whose square line element is given given by
$$
d{S}^2 = g_{\mu \nu}(\psi) dx^{\mu} dx^{\nu}.
$$
Consequently , applying the Euler Lagrange equation on following Lagrangian,
\begin{equation}
L= g_{\mu \nu}(\psi) U^{\mu}U^{\nu},
\end{equation}
  one obtains the geodesic equation,
 \begin{equation}
 \frac{d U^{\alpha}}{dS} +\Gamma^{\alpha}_{\beta \gamma}  U^{\beta}U^{\gamma} =0.
 \end{equation}
This equation can be transformed into INF, if one considers $x^{0}$  as one of parameters such that $x^{0} = c t$,
 \begin{equation}
\frac{d U^{\alpha}}{dS}= \frac{d U^{\alpha}}{dt} \frac{dt}{dS}
 \end{equation}
 and regarding $x^{0}$ the fourth component of the geodesic equation as
 \begin{equation}
 \frac{d U^{0}}{dS} =-\Gamma^{0}_{\beta \gamma}  U^{\beta}U^{\gamma}.
 \end{equation}
 Thus, substituting (94) and (95) in (93) and after some manipulations to get{\footnote{Verozub, private communication 2019}} [16]
 \begin{equation}
\frac{d^{2} x^{\alpha}}{d t^2} + ({\Gamma}^{\alpha}_{\beta \gamma}- c^{-1} {\Gamma}^{0}_{\beta \gamma}\frac{d x^{\alpha}}{dt} ) \frac{d x^{\beta}}{ds}\frac{d t^{\gamma}}{dt} =0
\end{equation}
which is exactly  obtained using Euler-Lagrange for the following Lagrangian of a point mass subject to a force field, defined in Minkowski space.
\begin{equation}
L = -mc (g_{\mu \nu} (\psi) \dot{x}^{\mu} \dot{x}^{\nu})^{\frac{1}{2}},
\end{equation}
Thus,  in order to extend type of  motion as described in (93) into of spinning objects, we must suggest the following Lagrangian functions. \\
{i.} Case $\bar{P}^{\alpha} = m \bar{U}^{\alpha}$ .\\
We suggest the  following Lagrangian to derive the spinning equation for short such that,
\begin{equation}
\bar{L}= G_{\alpha \beta}(\psi) \bar{U}^{\alpha} \frac{\bar{D} \bar{\Psi}^{\beta}}{\bar{D}S} + \bar{S}_{\mu \nu} \frac{\bar{D} \Psi^{\alpha \beta}}{ \bar{D}S} + \frac{1}{2m}\bar{R}_{\mu \nu \rho \delta} \bar{S}^{\rho \delta} \bar{U}^{\nu} \bar{\Psi}^{\mu}.
\end{equation}
Taking the variation with respect to $\bar{\Psi}^{\alpha}$ and $\bar{\Psi}^{\alpha \beta}$ to obtain
\begin{equation}
\frac{\bar{D \bar{U^{\alpha}}}}{\bar{D}\bar{S}} = \frac{1}{2m} \bar{R}^{\alpha}_{\beta \sigma \rho}\bar{S}^{\sigma \rho}\bar{U}^{\beta},
\end{equation}
and
\begin{equation}
\frac{\bar{D \bar{S^{\alpha \beta}}}}{\bar{D}\bar{S}} = 0.
\end{equation}
  In a similar way, using the commutation relations as shown in (11),(12), (13) and (14) , we obtain their corresponding deviation equations
\begin{equation}
\frac{\bar{D}^{2}\bar{\Psi}^{\alpha}}{\bar{S}^{2}}=  \bar{R}^{\alpha}_{\beta \sigma \rho}\bar{U}^{\beta} \bar{U}^{\sigma} \bar{\Psi}^{\rho} +\frac{1}{2m}( \bar{R}^{\alpha}_{\beta \sigma \rho}\bar{S}^{\sigma \rho}\bar{U}^{\beta}) _{; \rho} \bar{\Psi}^{\rho}
\end{equation}
and
\begin{equation}
\frac{\bar{D}^{2}\bar{\Psi}^{\alpha \beta}}{\bar{S}^{2}}=  \bar{S}^{[ \alpha \rho}R^{\beta ]}_{ \rho \sigma \delta}\bar{U}^{\sigma} \bar{\Psi}^{ \delta}
\end{equation}
{ii.} Case $\bar{P}^{\alpha}= mU^{\alpha} +U_{\beta} \frac{\bar{D} S^{\alpha \beta}}{\bar{D} S}$ \\
In this case, we suggest the corresponding the Bazanski-like Lagrangian for a spinning object to become,
\begin{equation}
\bar{L}= G_{\mu \nu}(\psi)  \bar{P}^{\mu}\frac{\bar{D} \bar{\Psi}^{\nu} }{\bar{D} \bar{S}} + \bar{S}_{\mu \nu}\frac{\bar{D} \bar{\Psi}^{\mu \nu} }{\bar{D} \bar{S}}  + \bar{f}_{\mu}\bar{\Psi}^{\mu} + \bar{f}_{\mu \nu}\bar{\Psi}^{\mu \nu}
\end{equation}
where,
$$
\bar{f}^{\alpha} = \frac{1}{2} \bar{R}^{\alpha}_{\beta \sigma \rho}\bar{S}^{\sigma \rho}\bar{U}^{\beta}
$$
and
$$
\bar{f}^{\alpha \beta} =   \bar{S}^{[ \alpha \rho} \bar{R}^{\beta ]}_{ \rho \sigma \delta}\bar{U}^{\sigma} \bar{\Psi}^{ \delta}  + 2 ( \bar{P}^{ [\alpha} \bar{U}^{\beta ]})_{; \delta} \Psi^{\delta}.
$$
and,taking the variation with respect to $\Psi^{\alpha}$ and $\Psi^{\alpha \beta}$ , therefore we obtain,
\begin{equation}
\frac{\bar{D \bar{P^{\alpha}}}}{\bar{D}\bar{S}} = \frac{1}{2} \bar{R}^{\alpha}_{\beta \sigma \rho}S^{\sigma \rho}U^{\beta},
\end{equation}
and
\begin{equation}
\frac{\bar{D \bar{S^{\alpha \beta}}}}{\bar{D}\bar{S}} = 2 \bar{P}^{[ \mu} \bar{U}^{\nu ]} ,
\end{equation}
While, in order to derive their counterparts deviation equations  following the same technique as mentioned in (11), (12) and (15), we get
\begin{equation}
\frac{\bar{D}^{2}\bar{\Psi}^{\alpha}}{\bar{D} \bar{S}^{2}}=  \bar{R}^{\alpha}_{\beta \sigma \rho}\bar{P}^{\beta} \bar{U}^{\sigma} \bar{\Psi}^{\rho} +\bar{f}^{\alpha} _{; \rho} \bar{\Psi}^{\rho}
\end{equation}
and
\begin{equation}
\frac{\bar{D}^{2}\bar{\Psi}^{\alpha \beta}}{\bar{D} \bar{S}^{2}}=  \bar{S}^{[ \alpha \rho}R^{\beta ]}_{ \rho \sigma \delta}\bar{U}^{\sigma} \bar{\Psi}^{ \delta}
\end{equation}
 From the above equations, we can figure out that, the spinning and spinning deviation equations as similar as the famous spinning and spinning equations [21] of general relativity. Such a similarity is due to replacing the metric tensor $g_{\mu \nu} (x)$ by $g_{\mu \nu} (\psi)$.
\section{Conclusions}
Equations of motion of spinning objects in different types of bi-metric theories of gravity have revealed the effect of different curvatures as shown(30), (42), (55) and (91) associated with the moving particles have been discussed. These curvatures are not the mere Riemanian curvature, due to the involvement of different factors affecting their appearance.

Accordingly, It has been shown that, in Rosen's theory that the  Papapetrou equation, for short, is expressed with one single curvature, while the other curvature is neglected due to its flatness of the associated space time as shown in (18) which appears its r.h.s. similar to (9); even with a combined affine derivatives.  Yet, the system of equations are quite different than their counterparts in general relativity . This difference is clarified on dealing with  $P^{\mu} = m U^{\mu} + U_{\nu} \frac{D S^{\mu \nu}}{D S}$   as shown  by comparing equations (9) and (10) with their counterparts in (23), (24), (36),(37),(47),(48),(63),(64),(65),(66), (79), (81),(100) and (101) . Also, we have found the effect of the other type of absolute covariant derivative, on the spinning deviating object, despite the vanishing of its curvatures as shown in equations (24), (25) and (26).

On displaying the set of spinning and spinning deviation equations of Moffat's version, as regarded to be a bi-metric theory of having instead of two separate metrics, one combined metric ,with an amended affine connection, stemmed from Wyel geometry [32] , having its own curvature (31), (32), (36) and (37) leading to similar appearance of the  Papapetrou equations and their deviation ones as in equations (33), (34), (36) and (37). By examining  Moffat's model, we have found the nessity to derive the system of equations of rotating objects. Such equations as expressed in Bi-metric theory of gravity  have the same appearance like the orginal Papapetrou equations for objects in  general relativity [19].\\

Since, the problem of bi-metric theory is assigned to define strong fields of gravity, therefore, it is worth mentioning to examine the bi-metric analog of MOND due to its role to find solutions to problems that have no general relativity explanation i.e. the rotation curves of spiral galaxies. This may give rise to consider, the derivation of sinning equations of BIMOND, has given rise, to investigate to what extend different metrics,  may affect on the behaviorism of spinning and spinning deviations equations as a result of finding a tensor connecting the two affine connections (40 ) has an impact on  the two  different types of curvature.  From this perspective, it was necessary to obtain their possible set of of equation to spinning equations (42), (43) , (47) , and (48)  and their deviation ones as in equations (44), (45), (49) and (50).

Nevertheless, the tendency of studying bi-metric theory has been developed by proposing  two different sources of gravity, by regarding bi-gravity theory of ghost-free [13] . This has inspired us to propose, such  a hypothetical definition of twin spin tensor, associated with twin matter. Owing to this illustration, we have figured out that there are two independent sets of spinning equations and their deviations ones stemmed from one Lagrangian. Such an approach may give rise to search for an appropriate geometry able to express these two types of matter together.

 Moreover, we have derived the set of spinning and spinning deviation equations for Verozub's version a bi-metric theory of gravity. In this type theories expressed the metric tensor is no longer a function of  coordinates  of space time $g_{\mu \nu}(x)$ ,  but a function based on a proposed field variable $(\psi)$. This may lead  to define a new metric tensor $G_{\mu \nu}$ its own affine connection $\bar{\Gamma}^{\gamma}_{\alpha \beta}$ and associated curvature $\bar{R}^{\alpha}_{\beta \rho \sigma} $. These quantities are playing a vital role for deriving their corresponding  spinning equations  and spinning deviation equations, different equations of spinning (95), (96), (100) and (101), as well as there a spinning deviation equations (97), (98), (102) and (103).

Finally, from the results obtained in our present work, it has become essential to search for an a wider geometry than the usual Riemannian one, able to  express all quantities of bi-gravity theory geometrically. This may be found by considering Finsler geometry, one of good candidates to fulfill this required task, which will be determined in our future work.\\
{\bf{Acknowledgement}}: The author would like to thank Professors L. Verozub  for his discussions.

\section*{References}

{[1]}N. Rosen  {\it{Ann. Physics}} ,{\bf{84}},455 (1974). \\
{[2] } A. Papapetrou   {\it{Proceedings of Royal Irish Academy Section A}}  {\bf{52}}, 11 (1948). \\
{[3]} N Rosen    {\it{Gen. Relativ.  and Gravit.}} {\bf{4}},  435 (1973). \\
{[4]}  M. Israelit  {\it{Gen. Relativ.  and Gravit.}} {\bf{7}},  623 (1976). \\
 {[5]} R. Falik, and N. Rosen  {\it{Gen. Relativ.  and Gravit.}} {\bf{13}},  599 (1981). \\
{[6]} J.W. Moffat  arXiv: hep-th/0208122 (2002) \\
{[7]} J.W.  Moffat   arXiv: quant-ph/0204151 (2002) \\
{[8]} M. Milgrom  {\it{Astrophys. J.}} {\bf{270}}, 365 (1983) \\
{[9]} M. Milgrom,  {\it{Phys.Rev.D80}} ,123536 (2009)\\
{[10]} M. Milgrom, arXiv: 1404.7661 (2014) \\
{[11]} M.  Milgrom,  {\it{Phys. Rev. D 89}}, 024027 (2014) \\
{[12]} S.,Hassan, and  R.A. Rosen   arXiv 1109.3515 (2012) \\
{[13]} E. Babichev, and R. Rito arXiv 1503.07529 (2015) \\
{([14]} Y. Akrami,  T. Kovisto, and  A.R. Solomon  arXiv.1404.0006 {(2014)}\\
 {[15]}  L.V. Verozub   arXiv 0911.5512 (2009)\\
{[16]} L.V. Verozub,   {{Space-time  Relativity and Gravitation, Lambert Academic Publishing.}}(2015)\\
{[17]} L.V. Verozub, and A. Kochetov  {\it{Astron. Nschr. }}{\bf{322}}, .(2001). \\
 {[18]} M.E. Kahil  {\it{Gravit. Cosmol.}}  {\bf{23}}, 70 (2017) \\
{[19]} A. Papapetrou  {\it{Proceedings of Royal Society London A}} {\bf{209}} , 248(1951)  \\
{[20]} M.E. Kahil {\it{Gravit. Cosmol.}}  {\bf{24}}, 83 (2018) \\
{[21]} M.E. Kahil {\it{ADAP}} {\bf{3}},  136 (2018) \\
{[22]}M.  Pavsic and M.E. Kahil Open Physics {\bf{10}}, 414 (2012). \\
{[23]} S.L. Bazanski  {\it{J. Math. Phys.}} {\bf {30}}, 1018 (1989) \\
{[24]} M.E. Kahil   {\it{J. Math. Phys.}} {\bf {47}},052501 (2006) \\
{[25]} M. Heydrai-Fard, M. Mohseni, and  H.R Sepanigi {\it{Phys. Lett. B}} {\bf{626}}, 230 (2005) \\
{[26]} M Roshan {\it{Phys.Rev. D}}{\bf{87}},044005 (2013) \\
{[27]}J Foukzoun, SA Podosenov, AA Potpov, and   E Menkova arXiv: 1007.3290 (2010)\\
{[28]}J.W. Moffat  	arXiv: 1110.1330 (2011) \\
{[29]}J.W. Moffat    arXiv: 1306.5470 (2013) \\
{[30]}S. Hossenfelder  arXiv: 0807.2838 (2008).\\
{[31]} JD Bekenstein  arXiv: gr-qc/9211017 (1992). \\
{[32]} C. Romero, J.B. Fonseca-Neto, and ML Pucheu  arXiv: 1201.1469 (2012). \\
\end{document}